\input head
\begin{document}
\begin{titlepage}
\begin{flushright}
Z\"urich University Preprint\\
ZU-TH 26/93
\end{flushright}
\vfill
\begin{center}
{\large\bf GRAVITATIONAL MICROLENSING: A METHOD
FOR DETECTING HALO DARK MATTER$^{\star}$ }\\
\vfill
\vskip 1.0cm
{\bf Philippe~Jetzer\footnote{Supported by the Swiss National
Science Foundation}}\\
\vskip 1.0cm
Institute of Theoretical Physics,
University of Z\"urich, Winterthurerstrasse 190,\\
CH-8057 Z\"urich, Switzerland
\end{center}
\vfill
\begin{center}
Abstract
\end{center}
\baselineskip=12pt
\begin{quote}
It has been shown by Paczy\'nski that gravitational
microlensing is potentially a useful method for detecting
the dark constituents of the halo of our galaxy, if
their mass lies in the approximate domain $10^{-6}<M/M_\odot<10^{-1}$.

Microlensing observations now under way monitor several millions of
stars in the Large Magellanic Cloud and in the Galactic Bulge.
Here I discuss the main features of the microlensing
events: in particular their rates and
probability, taking also into account a possible flattened shape for the halo.
\end{quote}
\vfill
\begin{center}
September 1993
\end{center}
$^{\star}$ To appear in the proceedings of
the workshop ``The dark side of the Universe: experimental
efforts and theoretical framework'',
Roma (Italy), June 1993.\\
\end{titlepage}
\newpage

\baselineskip=14pt
\noindent{\bf 1. INTRODUCTION}\\

One of the most important problems in today astrophysics, with possible
important implications for particle physics, lies in the knowledge of
the nature of the non-luminous matter of galactic halos.
The existence of the dark halo can be inferred
from the observed rotation curves.
The Milky Way and several other galaxies are observed
to be embedded in an ``invisible halo'' roughly an order of magnitude
more extended and more
massive than their visible domain, which consists
of stars, gas and dust.

The large amounts of unobserved dark matter may
be gathered in galactic halos in a
yet-undetected form, such as ``Massive Halo Objects'' (MHOs).
Except for super-massive black holes
($M_{BH}>100M_\odot$), MHOs,
if made of ordinary matter, should be bodies made of the lightest
primordial elements: Hydrogen and Helium. In the narrow range
$0.1>M/M_\odot>0.08$ they would be hard-to-spot
white dwarfs. Lighter MHOs,
below the nuclear-ignition threshold, would be ``brown
dwarfs'', Jupiter like bodies.
The possible origin of small hydrogenoid
planets, by fragmentation or by some other mechanism, is at present
not understood.  It has been pointed out that the mass distribution
of the MHOs, normalized to the dark halo mass density, could be
a smooth continuation of the known initial mass function
of ordinary stars$^1$.
The ambient radiation, or their own body heat, would make
sufficiently small objects made of H and He evaporate rapidly.
The condition that the rate of evaporation of such a hydrogenoid sphere be
insufficient to halve its mass in a billion years leads to the
following lower limit on their mass$^1$:
$M > 10^{-7} M_{\odot}
(T_S /30 K)^{3/2}$ ($T_S$ being their surface
temperature and $\rho \sim 1~ g~ cm^{-3}$ their average density).

Paczy\'nski$^2$
has shown how to
detect the MHOs if their mass lies in the range $10^{-6}
< M/M_{\odot} < 10^{-1}$ by means of the gravitational lensing effect.
Observations  which are now in progress by a French
collaboration$^{3,4,5}$
using the ESO
telescopes and  by a American-Australian collaboration$^6$
using the Mt. Stromlo Observatory
monitor several millions
stars in the Large Magellanic Cloud (LMC), whereas an American-Polish
collaboration working at Las Campanas Observatory operate on stars
located in the Galactic Bulge$^7$
.\\

\noindent{\bf 2. MICROLENSING PROBABILITY}\\

When a
MHO of mass $M$ is sufficiently close to the line between us and the
star, the light from the source suffers a gravitational
deflection. The deflection angle is usually so small that we do not see
two images but rather a magnification  of the original star brightness.
This magnification, at its maximum, is given by
\begin{equation}
A_{max}=\frac{u^2+2}{u(u^2+4)^{1/2}} . \label{eq:bb}
\end{equation}
Here $u=d/R_E$ ($d$ is the distance of the MHO from the line of sight)
and the Einstein radius $R_E$ is defined as
\begin{equation}
R_E^2=\frac{4GMD}{c^2}x(1-x) \label{eq:cc}
\end{equation}
with $x=s/D$, and
where $D$ and $s$ are the distance between the source, respectively the MHO and
the observer.

An important quantity is the optical depth $\tau_{opt}$
to gravitational microlensing defined as
\begin{equation}
\tau_{opt}=\int_0^1 dx \frac{4\pi G}{c^2}\rho(x) D^2 x(1-x)
\label{eq:za}
\end{equation}
with $\rho(x)$ the mass density of microlensing matter at distance
$s=xD$ from us. The quantity $\tau_{opt}$ is the probability
that a source is found within a radius $R_E$ of some MHO and thus has a
magnification that is larger
than $A_{max}= 1.34$ ($d \leq R_E$).

We calculate $\tau_{opt}$ for a galactic mass
distribution that has the form of
an axisymmetric spheroid, with the solar system in the symmetry plane.
In the reference system where the origin is at the galactic center (GC)
and the symmetry plane is the $x_1-x_2$ plane, we write the density at
$\vec r=(x_1,x_2,x_3)$ as
\begin{equation}
\rho(\vec r)=\frac{\rho_0(a^2+R^2_{GC})}
{a^2+x_3^2/q_H^2+x_1^2+x_2^2}.\label{eq:zb}
\end{equation}
Here $q_H$ is the axis ratio oblateness, $a$ is the core radius,
$\rho_0$ is the local dark mass
density in the solar system and $R_{GC}$ is the distance
between the observer and the $GC$. The local density of dark matter
depends on the galactic oblateness.
We have for example$^8$
that $\rho_0(q_H=0.5) \simeq 1.7 \rho_0(q_H=1)$.

When the source is so far that it can be considered out of our
galactic
halo, the upper limit of the integral in eq.(\ref{eq:za}) instead of 1
is $D_h/D$, where $D_h$ is the extent
of the galactic halo along the line of sight (l.o.s.).

The direction of the l.o.s. of the source is
determined by two angles. Let us denote by $\alpha$ the angle of the
l.o.s. with the direction of the GC and $\beta$ the angle
of the l.o.s. with the symmetry plane of the galactic spheroid. We find,
by performing the integral in eq.(\ref{eq:za}), that
\begin{equation}
\eqalign{\tau_{opt}=& \hat\tau_{opt}[\frac{-1}{C}
\frac{x_h}{x_s}+\frac{1}{C^2}(\frac{C}{2}-\frac
{cos\alpha}{x_s})ln(\frac{1+x_a^2-2x_h cos\alpha+Cx^2_h}{1+x_a^2})
+\frac{1}{C^2Bx_s}((1+x_a^2)C \cr &
-2 cos^2\alpha + Cx_s cos\alpha)(arctg \frac{Cx_h-cos\alpha}{B}+
arctg\frac{cos\alpha}{B})]} \label{eq:zc}
\end{equation}
with
\begin{equation}
C=\frac{1}{q_H^2}sin^2\beta+cos^2\beta, \label{eq:zca}
\end{equation}
\begin{equation}
B=[(1+x_a^2)C-cos^2\alpha]^{1/2}, \label{eq:zcb}
\end{equation}
\begin{equation}
\hat\tau_{opt}=\frac{4\pi G}{c^2}\rho_0 R_{GC}^2(1+x_a^2),
\label{eq:zcc}
\end{equation}
and $x_a=a/R_{GC}$, $x_s=D/R_{GC}$, $x_h=D_h/R_{GC}$.

In order to obtain the limit of the ``isothermal" sphere in the
former expression we have just to set $C=1$
and, with that, the dependence
on the angle $\beta$ disappears.
Standard values for the
parameters are$^9$
$\rho_0=0.3~GeV/cm^3=7.9~10^{-3} M_\odot/pc^3$,
$a=5.6~kpc$ and $R_{GC}=8.5~kpc$.
With these values we get, for a spherical halo, $\tau_{opt}=0.7~ 10^{-6}$
for the LMC and $\tau_{opt}=10^{-6}$ for the SMC, whereas for a nonspherical
halo with $q_H=0.5$: $\tau_{opt}=0.8~10^{-6}$ for the LMC and the
SMC$^{10}$
{}.

For globular clusters (with $q_H=1$) we get tipically
$\tau_{opt}\approx 4~ 10^{-8}$ (for instance:
NGC 288 $\tau_{opt}=6.3~ 10^{-8}$;
NGC 4833 $\tau_{opt}=5~ 10^{-8}$ and NGC 3201
$\tau_{opt}=2.8~ 10^{-8}$); compared to the
LMC or SMC this means a value $\approx$ 20 times smaller for the optical
depth $\tau_{opt}$.
However, should the planned observations of the
LMC show a lot of microlensing events, then the monitoring of
globular clusters could give, although with less data, useful
informations on the spatial distribution of the MHOs in the halo.

The magnification of the brightness of a star by a MHO is a time-dependent
effect (see Fig.1).
For a source that can be considered as
pointlike (this is the case if the projected star radius at the MHO
distance is much less than $R_E$)
the light curve as a function of time is obtained by inserting
\begin{equation}
u(t)=\frac{(d^2+v^2_Tt^2)^{1/2}}{R_E} \label{eq:zd}
\end{equation}
into eq.(\ref{eq:bb}),
where $v_T$ is the transverse velocity of the MHO, which can be inferred
from the measured rotation curves ($v_T \approx 200~ km/s$). The
achromaticity, symmetry and uniqueness of the signal are distinctive
features that will allow to discriminate a microlensing event from
background events such as variable stars.

The behaviour of the magnification with time, $A(t)$, determines two
observables namely, the magnification at the peak $A(0)$ - denoted
by $A_{max}$ -
and the width of the signal $T$, which we define as
being the time the microlensing signal has an increase in
magnitude greater than half the maximum magnitude increase.\\

\noindent{\bf 3. MAGNIFICATION FOR AN EXTENDED SOURCE}\\

A large fraction of the stars of a typical globular cluster are
red giants, as well as among the brighter stars in the Magellanic clouds.
For these stars the apparent radius projected on the deflector's plane:
$R_p=R_{star} x$, where $R_{star}$ is the radius of the star located
at a distance $D$, can be of the same size as the Einstein radius
$R_E$ of the lensing object. Therefore the point-source approximation
is no longer valid. Here we calculate the resulting magnification
assuming that the source has a uniform surface brightness. We follow the
derivation given by Maeder$^{11}$.

The distance between the center of the
disc of the star projected on the deflector's plane and the position of
the deflector at a given time $t$ is:
$\widehat d =\sqrt{d_0^2+(v_Tt)^2}$, where
$d_0$ is the minimal distance.
There are two cases to be considered:
(a) $\widehat d \leq R_p$ (see Fig.2a) and (b)
 $\widehat d >R_p$ (see Fig.2b). In case (a)
the surface of the image is delimited by the outer border-line
described by the vector of length $l$
and the inner border by the vector of
length $l^{\prime}$  with
\begin{equation}
l^2=\frac{b^2}{2}+R_E^2+\frac{b^2}{2}\sqrt{1+\frac{4R_E^2}{b^2}},
\label{eq:wn}
\end{equation}
\begin{equation}
l^{\prime 2}=\frac
{b^2}{2}+R_E^2-\frac{b^2}{2}\sqrt{1+\frac{4R_E^2}{b^2}},\label{eq:wo}
\end{equation}
and
\begin{equation}
b=\widehat d cos\beta+\sqrt{R_p^2-\widehat d ^2sin^2\beta}.\label{eq:wp}
\end{equation}
The magnification is given by the ratio between the surface of the
image and the surface of the disc of the star projected on the
deflector's plane
\begin{equation}
A=\frac{\int_0^{\pi}(l^2-l^{\prime 2})d\beta}{\pi R^2_p}.\label{eq:wq}
\end{equation}

In case (b), when the angle $\beta$ is going from $-\beta_0$ to
$\beta_0$, where $\beta_0=arcsin(R_p/d)$, $l(b_1)$ describes the surface
$S_1$ (see Fig.3) and $l(b_2)$ the surface $S_2$. Similarly
$l^{\prime}(b_1)$ describes  $S_1^{\prime}$ and $l^{\prime}(b_2)$
describes the surface $S_2^{\prime}$. The magnification is again given
by the ratio between the surface of the lensed image (=$S_1-S_2+
S_1^{\prime}-S_2^{\prime}$) and the surface of  the disc of the star
projected on the deflector's plane
\begin{equation}
A=\frac{1}{\pi R_p^2}\int_0^{\beta_0}(l_1^2-l_2^2)d\beta+\frac{1}
{\pi R_p^2}\int_0^{\beta_0}(l_1^{\prime 2}-l_1^{\prime 2})d\beta,
\label{eq:wr}
\end{equation}
where $l_1$,$l_2$
 stands for $l(b_1)$ and $l(b_2)$ (eq. (\ref{eq:wn})) and
similarly for $l_1^{\prime}$ and $l_2^{\prime}$ (eq.(\ref{eq:wo})) with
\begin{equation}
b_1=\widehat d cos\beta+\sqrt{R_p^2-\widehat d ^2sin^2\beta}\ ,
\label{eq:ws}
\end{equation}
\begin{equation}
b_2=\widehat d cos\beta-\sqrt{R_p^2-\widehat d ^2sin^2\beta}\ .
\label{eq:wt}
\end{equation}

The above eqs.(\ref{eq:wq}) and (\ref{eq:wr})
for the magnification can easily be integrated
numerically. The
maximum magnification is large provided that $R_p < R_E$. From this
condition one can estimate the minimum MHO mass below which the
magnification becomes negligible$^2$
:
\begin{equation}
M_{min}~\approx~1.2 \times 10^{-8}~M_{\odot}~
\left( \frac{s}{10~kpc}\right)
\left( \frac{R_{star}}{R_{\odot}}\right) ^2
\left( \frac{50~kpc}{D}\right) ^2\ , \label{eq:wnn}
\end{equation}
where $s$
is the distance of the MHO from the observer. From eq.(\ref{eq:wnn})
it follows that for a red giant with $R=200 R_{\odot}$ located in the LMC
the minimum lensing mass has to be
$\approx 0.5 \times 10^{-3}~M_{\odot}$.\\

\noindent{\bf 4. MICROLENSING RATES}\\

Microlensing rates depend on the mass and velocity distribution of
MHOs.
The mass density at a distance $s=xD$ from the observer is given by
eq.(\ref{eq:zb}), where we consider
from now on only the spherical symmetric
case ($q_H=1$).
The ``isothermal"
spherical halo model does not determine the MHO number density as a
function of mass. A
simplifying  assumption is to let the mass distribution be independent
of the position in the galactic halo, i.e., we assume the following
factorized form for the number density per unit mass $dn/dM$,
\begin{equation}
\frac{dn}{dM}dM=\frac{dn_0}{d\mu}
\frac{a^2+R_{GC}^2}{a^2+R_{GC}^2+D^2x^2-2DR_{GC}x cos\alpha}~d\mu=
\frac{dn_0}{d\mu} H(x) d\mu~,
\label{eq:zj}
\end{equation}
with $\mu=M/M_{\odot}$ ($\alpha$ is the angle of the l.o.s. with
the direction of the galactic center), $n_0$ not depending on $x$
and is subject to the normalization
$\int d\mu \frac{dn_0}{d\mu}M=\rho_0$.
Nothing a priori is known on the distribution $d n_0/dM$.

A different situation arises for the velocity
distribution in the ``isothermal"
spherical halo model, its
projection in the plane perpendicular to the l.o.s. leads to the following
distribution in the transverse velocity $v_T$
\begin{equation}
f(v_T)=\frac{2}{v_H^2}v_T e^{-v^2_T/v_H^2}.\label{eq:zr}
\end{equation}
($v_H \approx 210~km/s$ is the observed velocity dispersion in the halo).

In order to find the rate at which a single star
is microlensed with magnification
$A \geq A_{min}$, we consider MHOs
with masses between $M$ and $M+\delta M$, located at a distance from
the observer between $s$ and $s+\delta s$ and with transverse velocity
between $v_T$ and $v_T+\delta v_T$. The collision time can be
calculated using the well-known fact that the inverse of the collision
time is the product of the MHO number density, the microlensing
cross-section and the velocity.
The rate $d\Gamma$, taken also as a differential with respect
to the variable $u$, at which a single star is microlensed
in the interval $d\mu du dv_T dx$ is given
by$^{12,13}$
\begin{equation}
d\Gamma=2v_T f(v_T)D r_E [\mu x(1-x)]^{1/2} H(x)
\frac{d n_0}{d\mu}d\mu du dv_T dx,\label{eq:zt}
\end{equation}
with
\begin{equation}
r_E^2=\frac{4GM_{\odot}D}{c^2} \sim (21.7~ a.u.)^2 \sim
(3.2\times 10^9 km)^2 .\label{eq:zs}
\end{equation}

One has to integrate
the differential number of microlensing events,
$dN_{ev}=N_{\star} t_{obs} d\Gamma$,
over an appropriate range for $\mu$, $x$,
$u$ and $v_T$,
in order to obtain the total number of microlensing events which can
be compared with an experiment
monitoring $N_{\star}$ stars during an
observation time $t_{obs}$ and which is able to detect
a magnification such that $A_{max} \geq A_{TH}$.
The limits of the $u$ integration are determined by
the experimental threshold in magnitude shift, $\Delta m_{TH}$:
we have $0 \leq u \leq u_{TH}$.

The range of integration for $\mu$ is where the mass
distribution $dn_0/d\mu$ is not vanishing
and that for $x$ is
$0\leq x \leq D_h/D$ where $D_h$ is the extent of the galactic halo along
the l.o.s. (In the case of the LMC,
the star is inside the galactic halo and thus $D_h/D=1$.)
The galactic velocity distribution is cut at the escape velocity
$v_e \approx 640~km/s$ and therefore
$v_T$ ranges over $0\leq v_T \leq v_e$.
In order to simplify the integration we integrate $v_T$
over all the positive axis, due to the exponential factor in $f(v_T)$ the
so committed error is negligible.

However, the integration range of $d\mu du dv_T dx$
does not span all the interval we have just described.
Indeed, each experiment has time
thresholds $T_{min}$ and $T_{max}$ and only detects events with:
$T_{min}\leq T \leq T_{max}$,
and thus the integration range has to be such that $T$ lies in this
interval.
The total number of micro-lensing events is then given by
\begin{equation}
N_{ev}=\int dN_{ev}~\Theta (T-T_{min})\Theta (T_{max}-T),\label{eq:th}
\end{equation}
where the integration is over the full range of
$d\mu du dv_T dx$. $T$ is related in a complicated way
to the integration variables,
because of this, no direct
analytical integration in eq.(\ref{eq:th}) can be performed.

To evaluate eq.(\ref{eq:th}) we define
an efficiency function $\epsilon_0(\bar\mu)$
which measures the fraction of the total number of microlensing events
that meet the condition on $T$ at a
fixed MHO mass $M=\bar\mu M_{\odot}$.
A more detailed analysis$^{12}$
shows that
$\epsilon_0(\mu)$
is in very good approximation equal to unity for possible MHO objects
in the mass
range of interest here.
We now can write the total number of events in
eq.(\ref{eq:th}) as
\begin{equation}
N_{ev}=\int dN_{ev}~\epsilon_0(\mu).\label{eq:tl}
\end{equation}
Due to the fact that
$\epsilon_0$ is a function of $\mu$ alone, the integration in
$d\mu du dv_T dx$ factorizes into four integrals with independent
integration limits.

In order to quantify the expected number of events it is convenient
to take as an example a delta function distribution for the mass.
The rate of microlensing
events with
$A \geq A_{min}$ (or $u \leq u_{max}$), is then
\begin{equation}
\Gamma(A_{min})=\tilde\Gamma u_{max}=
D r_E u_{max} \sqrt{\pi}~v_H \frac{\rho_0}{M_{\odot}}\frac{1}{\sqrt{\bar \mu}}
\int^{D_h/D}_0 dx[x(1-x)]^{1/2} H(x)~.\label{eq:ta}
\end{equation}

Inserting the numerical values for the LMC
(D=50~kpc and $\alpha=82^0$) and taking $D_h/D=1$ we get
\begin{equation}
\tilde\Gamma=4
\times 10^{-13}~\frac{1}{s}~\left( \frac{v_H}{210~km/s}\right)
 \left(\frac{1}{\sqrt{D/kpc}}\right)
 \left( \frac{\rho_0}{0.3~GeV/cm^3}\right)
\frac{1}{\sqrt{M/M_{\odot}}}\ .
\label{eq:tb}
\end{equation}
For an experiment monitoring $N_{\star}$ stars during an
observation time $t_{obs}$ the total number of events with a
magnification $A \geq A_{min}$ is:
$N_{ev}(A_{min})=N_{\star} t_{obs} \Gamma(A_{min})$.
In the following table we show some values of $N_{ev}$ for the LMC,
taking
$t_{obs}=10^7$ sec ($\sim$ 4 Months), $N_{\star}=10^6$ stars and
$A_{min} = 1.34$ (or $\Delta m_{min} = 0.32$).

\begin{center}
\begin{tabular}{|c|c|c|c|c|}\hline
MHO mass in units of $M_{\odot}$ & Mean $R_E$ in km & Mean microlensing time &
$N_{ev}$ & $\epsilon_0$\\
\hline
$10^{-2}$ & $10^8$ & 9 days & 5 & 1 \\
$10^{-4}$ & $10^7$ & 1 day & 55 & 1 \\
$10^{-6}$ & $10^6$ & 2 hours & 554 & 1 \\
$10^{-8}$ & $10^5$ & 12 min. & 5548 & 0.6 \\
\hline
\end{tabular}
\end{center}

Gravitational microlensing could also be useful for detecting MHOs in
the halo of nearby galaxies$^{14,15}$
such as M31 or M33. In fact, it turns out
that the massive dark halo of M31 has an optical depth to microlensing
which is of about the same order of magnitude as that of our own
galaxy$^{14,16}$
$\sim 10^{-6}$. Moreover, an experiment monitoring stars in
M31 would be sensitive to both MHOs in our halo and in the one of M31.
One can also compute the microlensing rate$^{16}$
for MHOs in the halo of M31, for which we get
\begin{equation}
\tilde\Gamma=1.8 \times 10^{-12} \frac{1}{s} \left(\frac{v_H}{210~km/s} \right)
\left(\frac{1}{\sqrt{D/kpc}}\right)\left(\frac{\rho(0)}{1~Gev/cm^3} \right)
\frac{1}{\sqrt{M/M_{\odot}}}~. \label{eq:tc}
\end{equation}
($\rho(0)$ is the central density of dark matter.)
In the following table we show some values of $N^a_{ev}$ due to MHOs in the
halo of M31 with $t_{obs}= 10^7$ sec and $N_{\star}=10^6$ stars. In the
last column we give the corresponding number of events due to MHOs in our
own halo. The mean microlensing time is about the same for both types of
events.

\begin{center}
\begin{tabular}{|c|c|c|c|c|}\hline
MHO mass in units of $M_{\odot}$ & Mean $R_E$ in km & Mean microlensing time &
$N^a_{ev}$ & $N_{ev}$\\
\hline
$10^{-1}$ & $7\times 10^8$ & 38 days & 2 & 1 \\
$10^{-2}$ & $2\times 10^8$ & 12 days & 7 & 4 \\
$10^{-4}$ & $2\times 10^7$ & 30 hours & 70 & 43 \\
$10^{-6}$ & $2\times 10^6$ & 3 hours & 700 & 430 \\
\hline
\end{tabular}
\end{center}

$N_{ev}$ is almost by a factor of two bigger than $N_{ev}$. Of course these
numbers should be taken as an estimate, since they depend on the details
of the model one adopts for the distribution of the dark matter in the halo.\\

\noindent{\bf 5. MASS MOMENTS}\\

{}From the
experimental lensing data it is possible to extract informations on the
MHO mass distribution$^{12}$
. It
proves most useful to average powers of the width $T$, i.e., to construct
width moments.

Consider an experiment that observes $N_{\star}$
stars during a time
$t_{obs}$, that has thresholds $T_{min}, T_{max}$ and $\Delta m_{TH}$,
and that has recorded a set of microlensing events, each one with a
width $T$ and magnification at the peak $A_{max}$. For each event we get
a value for the dimensionless variable $\tau$, defined as follows
\begin{equation}
\tau = \frac{v_H}{r_E} \frac{T}{\sqrt{8g(A_{max})}}=\frac{v_H}{v_T}
[\mu x(1-x)]^{1/2}
{}~~~with ~~~g(A)=\frac{1}{\sqrt{1-1/A}}-\frac{1}{\sqrt{1-1/A^2}}~,
\label{eq:ts}
\end{equation}
and  we construct the
$n$-moment of $\tau$ from experimental data as:
$< \tau^n >= \sum_{events} \tau^n$.

The theoretical
expectations for these moments, which are given by
\begin{equation}
< \tau^n >=\int dN_{ev}~\epsilon_n(\mu)~\tau^n,\label{eq:wb}
\end{equation}
can be calculated and lead to
\begin{equation}
<\! \tau^n \!> \, = V \,u_{TH}\,\,
 \Gamma(2-m)\, \widehat H(m) <\!\mu^m\!>,
\label{eq:wd}
\end{equation}
with $m \equiv (n+1)/2$ and
\begin{equation}
V \equiv 2N_\star t_{obs} D \, r_E \, v_H
  = 26\,{\rm pc}^3\,\, \frac {N_\star}{10^6}
  \, \frac {t_{obs}}{4\,{\rm months}},\label{eq:we}
\end{equation}
\begin{equation}
\Gamma(2-m) \equiv \int_0^\infty \left[\frac{v_T}{v_H}\right]^{1-n}\,
      f(v_T)\,dv_T,      \label{eq:wf}
\end{equation}
\begin{equation}
\widehat H(m) \equiv \int^1_0 [x(1-x)]^m H(x)\, dx,\label{eq:wg}
\end{equation}
\begin{equation}
<\! \mu^m \!> \, \equiv \int d\mu~\epsilon_n(\mu)
\frac{dn_0}{d\mu} \mu^m.\label{eq:wh}
\end{equation}
Notice that
the convergence of the integrals in $v_T$ and $x$ restricts the
moments to the range $-3< n <3~(-1< m < 2)$.

The fundamental point of eq.(\ref{eq:wd}) is that the
experimental value of $<\tau^n>$,
with $n=2m-1$, allows us to determine the
moments of the MHO mass distribution, which can be used to reconstruct the
mass function itself.

We may use
$< \mu^1 >$ to know the fraction $f$ of the dark matter density $\rho_0$
that has been detected. Indeed, we have
\begin{equation}
f=\frac{M_J}{\rho_0}< \mu^1 > \approx 0.12~ pc^3 < \mu^1 > ,\label{eq:wl}
\end{equation}
and we expect of course $f \leq 1$. The fraction $f$
may be smaller than 1
because MHO's do not constitute all the galactic dark matter or/and
there are MHO's that have escaped detection because of the experimental
limitations.\\


\begin{figcap}
\item  Plot of
$-\Delta m=2.5~log A(t)$ as a function of time (in units of $10^5$
sec) for a pointlike source at the distance of the LMC.  The MHO is at
a distance of 25 kpc and has a transverse velocity of $200~km/s$ and a
mass of $10^{-2}~M_{\odot}$.  The higher curve, that serves to define
the event duration $T$, corresponds to $d=1/2~R_E$, whereas the lower
one has $d=R_E$.

\item The circle of radius $R_p$ centered at S is the unmodified
projected image
of the source star in the plane perpendicular to the line of sight and
located at the deflector's position. The hatched area is the observed
image of the star modified by the presence of a deflector centered at
D. $\widehat d$ is the projected distance between S and D.
a) $\widehat d \leq R_p$.
b) $\widehat d > R_p$.

\item Definition of surfaces of integration.
\end{figcap}
\end{document}